\documentclass[12pt,sd,amsmath,amssymb,reprint,superscriptaddress,floatfix,preprintnumbers]{iopart}
\usepackage{graphicx, multirow}
\usepackage{dcolumn}
\usepackage{bm}

\usepackage[dvipsnames]{xcolor}
\usepackage{epstopdf}
\usepackage{hyperref}
\usepackage{iopams}
\usepackage{framed}
\usepackage{inputenc} 
\usepackage{soul}

\expandafter\let\csname equation*\endcsname\relax
\expandafter\let\csname endequation*\endcsname\relax
\usepackage{amsmath,amsfonts,amssymb,cancel}

\date{}

\begin{document}
\title{Simplicial SIS model in scale-free uniform hypergraph}
\author{Bukyoung Jhun, Minjae Jo, and B. Kahng$^*$}
\address{$^{1}$CCSS, CTP and Department of Physics and Astronomy, Seoul National University, Seoul 08826, Korea }
\ead{$^*$bkahng@snu.ac.kr}
\vspace{10pt}

\begin{abstract}
The hypergraph offers a platform to study structural properties emerging from more complicated and higher-order than pairwise interactions among constituents and dynamical behavior such as the spread of information or disease. Recently, a simplicial contagion problem was introduced  and considered using a simplicial susceptible-infected-susceptible (SIS) model. Although recent studies have investigated random hypergraphs with a Poisson-type facet degree distribution, hypergraphs in the real world can have a power-law type of facet degree distribution. Here, we consider the SIS contagion problem on scale-free uniform hypergraphs and find that a continuous or hybrid epidemic transition occurs when the hub effect is dominant or weak, respectively. We determine the critical exponents analytically and numerically. We discuss the underlying mechanism of the hybrid epidemic transition. 
\end{abstract}


\section{Introduction}

In past decades, extensive research has been done on emerging phenomena in complex networks, including the spread of epidemic diseases and innovations~\cite{katona2011, rogers2004, rogers2005, xu2008}, opinion formation~\cite{acemoglu2013, grabowski2006, watts2007}, and many other topics~\cite{borgatti2003, boccaletti2006, hoang2003, newamn2003, provan2007}. An important issue for such emerging phenomena is to understand the origin and properties of phase transitions. Complex networks represented by graphs enable researchers to study such issues successfully. A graph is a collection of vertices and edges, where an edge represents a pairwise interaction between two vertices. In complex systems, however, interactions among constituents can be more complex than pairwise. For instance, more than two people can collaborate on a team. 

A {\sl hypergraph} is a generalization of a graph whose hyperedge connects two or more vertices. Consequently, it can be used to encode complicated social interactions that the graph representation cannot. In this hypergraph representation, a hyperedge of size $n$ connects $n$ researchers who collaborate on one task, for instance, $d$ authors of a $d$-author paper in coauthorship networks~\cite{xie2016}. This hypergraph representation successfully accounts for various types of collaborations~\cite{ghoshal2009, bolle2006, bolle2008, klamt2009, taramasco2010, vazquez2008, zhang2010, zlatic2009}. In particular, a {\sl uniform hypergraph} is one in which all the hyperedges have the same size. If the size of these hyperedges is $d$, the structure is called a $d$-uniform hypergraph, or $d$-hypergraph. Uniform hypergraphs can describe systems in which a uniform number of agents interact at the same time. Trivially, a 2-uniform hypergraph reduces to a graph. Owing to its simplicity, the uniform hypergraph enables succinct expression of diverse static and dynamic problems in terms of linear algebra using the adjacency tensor~\cite{courtney2016}. 

{\sl A simplicial complex} is a particular hypergraph with an additional constraint: If a hyperedge is in a simplicial complex, any non-empty subset of vertices in the hyperedge is also a hyperedge of the simplicial complex. This requirement makes the simplicial complex an appropriate tool for studying systems with high-order interactions, i.e., interactions that involve a large number of agents, which also include lower-order interactions. A hyperedge in a simplicial complex is often called a simplex. The simplicial complex has been a topic of extensive research. Examples include the collaboration network~\cite{ciftcioglu2017, patania2017}, semantic network~\cite{sizemore2017}, cellular network~\cite{estrada2018}, and brain network~\cite{petri2014, lee2012}.

{\sl A simplicial contagion} model was recently introduced~\cite{iacopini2019} to describe a complex contagion process on simplicial complexes; however, the model can also be easily applied to general hypergraphs. Here, we consider this simplicial contagion process on $d$-uniform hypergraphs with hyperedges of the same size. Specifically, we consider the case that infection spreads only when all but one of the nodes in the hyperedge are infected. Even though this is a simple case with a maximally conservative contagion process, it provides an essential factor that leads to a hybrid epidemic transition on hypergraphs. Here, we consider a simplicial susceptible-infected-susceptible ($s$-SIS) model, where infection spreads by a simplicial contagion process. Each node is in either the susceptible ($S$) or infected ($I$) state. A susceptible node becomes infected at a rate $\beta$ when all the other nodes in the same hyperedge are infected. If a node is infected, it changes spontaneously to the susceptible state $S$ at a rate $\mu$. This recovery process ($I \to S$) is defined as in the SIS model of a network because the recovery process occurs on each node independently, making it irrelevant to the structural type of the contagion process. 

Here we explore the $s$-SIS model on scale-free (SF) uniform hypergraphs. We use the annealed approximation for the static model of the uniform hypergraph, which is extended from the static model of the complex graph~\cite{goh2001b}. We find analytically that there exists a characteristic degree $\lambda_c=2+1/(d-1)$ such that when the exponent $\lambda$ of the degree distribution is $2 < \lambda \le \lambda_c$, a continuous transition occurs; however, when $\lambda > \lambda_c$, a hybrid phase transition occurs. In this hybrid phase transition, the order parameter jumps at a macroscopic scale and then increases continuously with criticality as a control parameter, $\eta\equiv \beta/\mu$, is increased. 

\section{Static model of uniform hypergraph \label{sec:static_model}}

The static model of a complex network~\cite{goh2001b,lee2006a} has been widely used to generate SF networks owing to its simplicity and analytical tractability. The model has been used to study the $q$-state Potts model~\cite{lee2005}, sandpile model~\cite{goh2003}, spin glasses~\cite{kim2005}, and many other topics~\cite{ghim2004,yang2008,lee2006b,yi2008} involving complex networks. 

A static model of a uniform hypergraph is a generalization of the static model of a complex graph. 
The static model of a $d$-uniform hypergraph is generated as follows:
\begin{enumerate}
\item[i)] Set the number of nodes in the system, $N$. 
\item[ii)] Assign each node a weight $p_i$ as 
\begin{equation}
p_{i}=\frac{i^{-\mu}}{\zeta_{N}(\mu)}\simeq\frac{1-\mu}{N^{1-\mu}}i^{-\mu},
\label{eq:probability_node}
\end{equation}
where $\zeta_{N}(\mu)=\sum_{j=1}^{N}j^{-\mu}$, and $0<\mu<1$. The normalization condition $\sum_{i=1}^{N}p_{i}=1$ is satisfied. 
\item[iii)] Select $d$ distinct nodes with probabilities $p_{i_1}\cdots p_{i_d}$. If the hypergraph does not already contain a hyperedge of the chosen $d$ nodes, then add the hyperedge to the hypergraph.
\item[iv)] Repeat step iii) $NK$ times. 
\end{enumerate}
Then, each node $i$ has average degree $\langle k_i \rangle$. These average degrees have a power-law distribution $P_d(k)\sim k^{-\lambda}$ with $\lambda=1+1/\mu$, where the brackets of $\langle k_i \rangle$ are omitted. The details are presented in \ref{appendix:degree_dist}. The minimum degree is obtained as $k_{\rm min}={N^{1-\mu}\langle k \rangle }/{\sum_{j=1}^{N}j^{-\mu}}$, which converges to a finite value, $\frac{\lambda-2}{\lambda-1}\langle k \rangle$, where $\langle k \rangle$ denotes the mean degree $\sum_k kP_d(k)$. The maximum degree is obtained as $k_{\rm max}={N\langle k \rangle}/{\sum_{j=1}^{N}j^{-\mu}}$, which behaves as $\frac{\lambda-2}{\lambda-1}\langle k \rangle N^{1/(\lambda-1)}\sim N^{1/(\lambda-1)}$. Thus, it diverges as $N\to \infty$. Hereafter, the minimum degree is denoted as $k_m$. Throughout this algorithm, $NK$ hyperedges are generated. 

\begin{figure}
\includegraphics[width=1\textwidth]{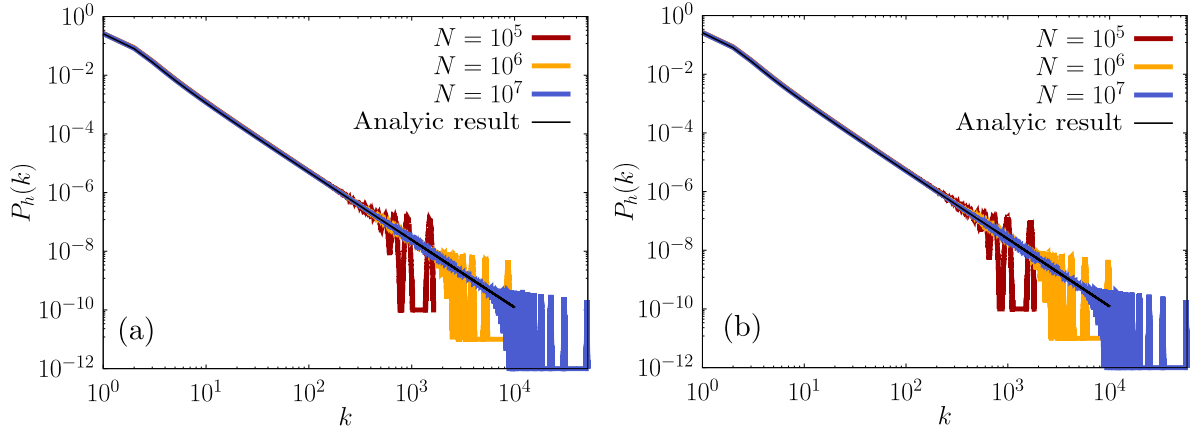}
\caption{Degree distribution of the static model of (a) $2$-uniform (graph) and (b) $3$-uniform hypergraph generated with the fitness exponent $1/\mu=1.3$. The system size $N$ is given as $N=10^5, 10^6,$ and $10^7$. As the system size is increased, the tail part of the degree distribution is extended, and power-law behavior with exponent $\lambda =1+{1}/{\mu}=2.3$ is confirmed.}
\end{figure} 

The probability that a hyperedge composed of $d$ distinct nodes $\{i_1 \cdots i_d \}$ is present is given by 
\begin{equation}
f_{i_{1}\cdots i_{d}}=1-\left(1-d! p_{i_{1}}\cdots p_{i_{d}}\right)^{NK}\simeq1-e^{-d! NKp_{i_{1}}\cdots p_{i_{d}}},
\end{equation}
and the probability that a hypergraph $G$ is generated is
\begin{equation}
P(G)=\prod_{a_{i_{1}\cdots i_{d}}\in G}\left(1-e^{-d! NKp_{i_{1}}\cdots p_{i_{d}}}\right)\prod_{a_{i_{1}\cdots i_{d}}\notin G}e^{-d! NKp_{i_{1}}\cdots p_{i_{d}}}.
\end{equation}
Because $d! NKp_{i_{1}}\cdots p_{i_{d}}\sim N^{d\mu-d+1}/\left(i_{1}\cdots i_{d}\right)^{\mu}$, for $0<\mu<\frac{d-1}{d}$, which is equivalent to $\lambda>2+\frac{1}{d-1}$, 
\begin{equation}
f_{i_{1}\cdots i_{d}}\simeq d! NKp_{i_{1}}\cdots p_{i_{d}},
\end{equation}
and for $2<\lambda<2+\frac{1}{d-1}$, 
\begin{equation}
f_{i_{1}\cdots i_{d}}\simeq\begin{cases}
1 & \left(i_{1}\cdots i_{d}\right)^{\mu}\ll N^{d\mu-d+1}\\
d! NKp_{i_{1}}\cdots p_{i_{d}} & \left(i_{1}\cdots i_{d}\right)^{\mu}\gg N^{d\mu-d+1}
\end{cases}.\label{eq:f_cases}
\end{equation}
We note that $\lambda=2+1/(d-1)$ is a characteristic degree and is denoted as $\lambda_c$, which reduces to $\lambda_c=3$ for an SF graph ($d=2$) and $\lambda_c < 3$ for an SF hypergraph ($d > 2$). The fraction of nodes that satisfies the second case of Eq.~\eqref{eq:f_cases} is proportional to $1-AN^{d\mu-d}$, where $A$ is a constant, which converges to one as $N\rightarrow\infty$. For $d=2$, the static model of the hypergraph reduces to the static model of the graph. For $\mu=0$, i.e., $\lambda=\infty$, the expected degree of all the nodes is identical, and the model reduces to an Erd{\H o}s--Renyi-like hypergraph.    

\section{Simplicial SIS model\label{sec:simplicial_sis_model}}

\begin{figure}
\includegraphics[width=1\textwidth]{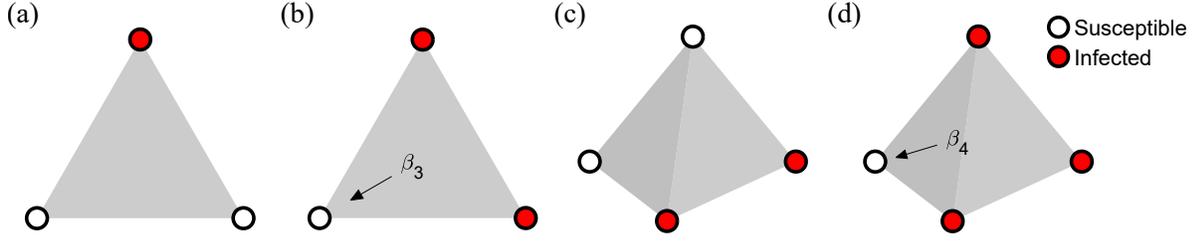}
\caption{Schematic illustration of the simplicial contagion process through hyperedges of size 3 in (a) and (b), and 4 in (c) and (d). The susceptible and infected nodes are depicted as white open circles and red filled circles, respectively. When $d-1$ of $d$ nodes in a hyperedge are infected, the infection spreads to the remaining susceptible node through the hyperedge at a rate $\beta_{d}$.}
\label{def_scp}
\end{figure}

A contagion process through an edge on a graph is called a simple contagion process. Simple contagion processes on complex graphs have been extensively studied to describe the spread of disease~\cite{nsoesie2014,tizzoni2012}, adoption of innovation~\cite{valente1996,rogers2005}, and opinion formation~\cite{acemoglu2013,grabowski2006,watts2007}. However, social phenomena that cannot be reduced to simple contagion processes have been observed, for instance, belief in bizarre urban legends~\cite{heath2001}, adoption of unproven new technologies~\cite{coleman1966}, willingness to participate in risky migrations~\cite{macdonald1964}, and the appeal of avant-garde fashion~\cite{crane1999}, and they depend on contact with multiple early adopters. Adoption of behaviors that are costly, risky, or controversial often requires affirmation or reinforcement from an independent source. More complicated models of contagion, namely, a complex contagion process, have been proposed to describe such social phenomena. Examples include the threshold model~\cite{granovetter1978,watts2002} and a generalized epidemic model~\cite{janssen2004,choi2017a}. 

A recently introduced simplicial contagion model~\cite{iacopini2019} represents a complex contagion process on a hypergraph. It applies a maximally conservative contagion process on the hypergraph, in which contagion through a hyperedge of size $d$ occurs only when all but one of the nodes in the hyperedge are infected. When this condition is met, the remaining susceptible node is infected at a rate $\beta_d$ per unit time. For instance, when nodes $j$ and $k$ are infected in the hyperedge $\{i,j,k\}$, node $i$ is infected with probability $\beta_3 \delta t$ in duration $\delta t$. If only node $j$ is infected and the other node, $k$, is not, the infection does not spread to node $i$ through the hyperedge. 

The complex contagion process in a $d$-uniform hypergraph is described by an adjacency tensor of dimension $d$. The rate equation is written as follows:
\begin{equation}
\frac{d}{dt}q_{i_1}=-\mu q_{i_1}+\frac{1}{\left(d-1\right)!}\left(1-q_{i_1}\right)\beta_{d}\sum_{i_{2}\cdots i_{d}}a_{i_{1}\cdots i_{d}}q_{i_{2}}\cdots q_{i_{d}},
\end{equation}
where $q_{i_1}$ is the probability that a node $i_1$ is infected, and $a_{i_{1}\cdots i_{d}}$ is the adjacency tensor, where $a_{i_{1}\cdots i_{d}}=1$ if nodes $\{i_{1}\cdots i_{d}\}$ are fully connected, and otherwise, it is zero. 

\section{Heterogeneous mean-field theory (annealed approximation)}

We use the heterogeneous mean-field theory to study the stationary states of the SIS model on SF $d$-uniform hypergraphs. This theoretical approach has been successful for examining the SIS~\cite{pastor-satorras2001a,pastor-satorras2001b} and susceptible-infected-recovered~\cite{moreno2002} models on SF graphs. It represents well the significant effect of a small portion of nodes with large degrees. Here, we consider the SIS model on SF $d$-dimensional uniform hypergraphs. We set up a differential equation for the density of infected nodes of degree $k$ and then obtain the self-consistency equation for the stationary solution. We solve a self-consistency equation to calculate the density of infected nodes as a function of infection rate. We investigate the properties of the epidemic transition. 

\subsection{Self-consistency equation}

The density of infected nodes with degree $k$, denoted as $\rho_k$, evolves with time as follows:
\begin{equation}
\frac{d}{dt}\rho_{k}=-\mu\rho_{k}+\beta \left(1-\rho_{k}\right)k\Theta^{d-1}.
\end{equation}
The first term on the r.h.s.~of the above equation is a loss term associated with the recovery process $I\to S$. The second term is a gain term associated with the contagion process $(d-1)I+S\to dI$. That is, a given node $i$ in state $S$ is changed to state $I$ by contagion from $d-1$ infected nodes in a hyperedge of size $d$ at a rate $\beta$, which is equivalent to $\beta_d$ in the previous notation, in which node $i$ is included. $\Theta$ is given by 
\begin{equation}
\Theta=\frac{\sum_{k=k_{m}}^{\infty}kP_h(k)\rho_{k}(t)}{\langle k \rangle},\label{eq:theta}
\end{equation}
where $kP_h(k)\rho_k/\langle k \rangle$ is the probability that a node connected to a randomly chosen hyperedge has degree $k$ and is infected at time $t$. We are interested in the behavior of $\rho_k$ in the stationary state, in which $d\rho_k/dt=0$, and we set $\eta \equiv \beta/\mu$ for convenience. 

The stationary solution of $\rho_{k}$ is obtained as 
\begin{equation}
\rho_{k}=\frac{\eta k\Theta^{d-1}}{1+\eta k\Theta^{d-1}}.\label{eq:rho_k}
\end{equation}
This solution implies that the infection probability $\rho_k$ always increases and approaches one as $k\rightarrow\infty$ for $\eta > 0$, and that it is controlled by a single factor, $\eta \Theta^{d-1}$. The density of infected nodes becomes $\rho \equiv \sum_k P_h(k)\rho_{k}$, which serves as the order parameter of the epidemic transition. 

To obtain $\rho$, we set up a self-consistency equation for $\Theta$ in the stationary state as follows:
\begin{equation}
\Theta=\frac{1}{\langle k\rangle }\sum_{k}kP_h(k)\rho_{k}=\frac{1}{\langle k\rangle }\sum_{k}kP_h(k)\frac{\eta k\Theta^{d-1}}{1+\eta k\Theta^{d-1}}.
\end{equation}
We define the self-consistency function $G(\Theta)$ as
\begin{equation}
G(\Theta)=\frac{1}{\langle k \rangle }\sum_{k}kP_h(k)\frac{\eta k\Theta^{d-1}}{1+\eta k\Theta^{d-1}}-\Theta
\end{equation}
and then obtain a solution $\Theta_0$ of $G(\Theta_{0})=0$. 

For the power-law degree distribution, $P_h(k)={(\lambda-1)}{k_m^{\lambda-1}}k^{-\lambda}$ for $k\geq k_m$, and the mean degree $\langle k \rangle=\frac{\lambda-1}{\lambda-2}k_m$, 
\begin{equation}
G(\Theta)=(\lambda-2)k_m^{\lambda-2}\sum_{k}k^{1-\lambda}\frac{\eta k\Theta^{d-1}}{1+\eta k\Theta^{d-1}}-\Theta.\\
\end{equation}
We treat $k$ as a continuous variable and recast the summation $\sum_{k=k_m}^{\infty}\cdots$ as the integration $\int_{k_m}^{\infty}dk \cdots$. 
\begin{eqnarray}
G(\Theta)&=&(\lambda-2)k_m^{\lambda-2}\int_{k_m}^{\infty}dkk^{-\lambda+1}\big(1+\frac{1}{ \eta k\Theta^{d-1}}\big)^{-1}-\Theta \\ 
&=&(\lambda-2)\int_{0}^{1}dzz^{\lambda-3}\big(1+\frac{z}{\eta k_m\Theta^{d-1}}\big)^{-1}-\Theta \label{eq:14}\\ 
&=&{}_{2}F_{1}\left(\lambda-2,1; \lambda-1;-\frac{1}{\eta k_m \Theta^{d-1}}\right)-\Theta, \label{eq:15}
\end{eqnarray}
where we changed the variable $k$ to $z$ as $z=k_m/k$ in Eq.~\eqref{eq:14}, and ${}_{2}F_{1}(a,b;c,d)$ in Eq.~\eqref{eq:15} is the Gauss hypergeometric function, which is defined as~\cite{abramowitz1965}
\begin{equation}
{}_{2}F_{1}(a,b;c,z)=\frac{\Gamma(c)}{\Gamma(b)\Gamma(c-b)}\int_0^1 dz z^{b-1}(1-z)^{c-b-1}(1-tz)^{-a}.
\end{equation} 

To obtain a solution $\Theta_0$ of $G(\Theta_{0})=0$, we first note that the self-consistency function has the following properties: $G(0) = 0$, and $G(1) < 0$. Second, we examine the derivative with respect to $\Theta$, which can be written as 
\begin{equation}
G^{\prime}(\Theta)=\frac{(d-1)(\lambda-2)}{k_m\eta\Theta^{d}(\lambda-1)}{}_{2}F_{1}\left(\lambda-1,2;\lambda;-\frac{1}{k_m\eta\Theta^{d-1}}\right)-1.
\label{eq:17}
\end{equation}
If $\lim_{\Theta \rightarrow 0} G^{\prime}(\Theta) > 0$, there exists at least one nonzero solution $\Theta_0$. Using the asymptotic properties of the hypergeometric function, we find that there exists a characteristic degree exponent $\lambda_c=2+1/(d-1)$ such that  
\begin{equation}
\lim_{\Theta\rightarrow0}G^{\prime}(\Theta)=\begin{cases}
+\infty & {\rm for}~~~ \lambda< \lambda_c\\
\frac{\pi/(d-1)}{\sin\left(\pi/(d-1)\right)}\left(k_m\eta\right)^{1/(d-1)}-1 & {\rm for}~~~\lambda=\lambda_c\\
-1 & {\rm for}~~~\lambda> \lambda_c
\end{cases}\label{eq:gtheta_derivative_asymptotic}.
\end{equation}
See \ref{appendix:gtheta_derivative} for details.

After we obtain $\Theta_0$, the density of infection $\rho$, which serves as the order parameter for the epidemic transition, is calculated as follows: 
\begin{equation}
\rho =\int_{k_m}^{\infty}dkP_h(k)\frac{\eta k\Theta_0^{d-1}}{1+\eta k\Theta_0^{d-1}}={}_{2}F_{1}\left(\lambda-1,1;\lambda;-\frac{1}{k_m\eta\Theta_0^{d-1}}\right)\label{eq:rho_hypergeom}.
\end{equation}
 We will determine the solution $\Theta_0$ and $\rho$ for each case in Eq.~\eqref{eq:gtheta_derivative_asymptotic} in the next section. 

\section{Phase transition and critical behavior}
\label{sec:analyticalResults}
The type of phase transition and the epidemic threshold are determined by the behavior of $G(\Theta)$, which in turn is determined by $\lim_{\Theta\rightarrow0}G^{\prime}(\Theta)$. Accordingly, we consider the epidemic transition separately for each case in Eq.~\eqref{eq:gtheta_derivative_asymptotic}. 

\subsection{Order parameter}

To solve Eqs.~\eqref{eq:15} and~\eqref{eq:rho_hypergeom}, we use a Taylor expansion of the hypergeometric function
\begin{align}
{}_{2}F_{1}\left(\lambda-2,1;\lambda-1;-\frac{1}{k_{m}\eta\Theta^{d-1}}\right)&=
\frac{(\lambda-2)\pi}{\sin(\pi\lambda)}(k_{m}\eta\Theta^{d-1})^{\lambda-2}\nonumber\\&+(\lambda-2)\sum_{n=1}^{\infty}(-1)^n\frac{(k_{m}\eta\Theta^{d-1})^n}{n-(\lambda-2)}\,.
\label{eq:hypergeometric_taylor}
\end{align}

\begin{figure}
\begin{center}
\includegraphics[width=0.5\textwidth]{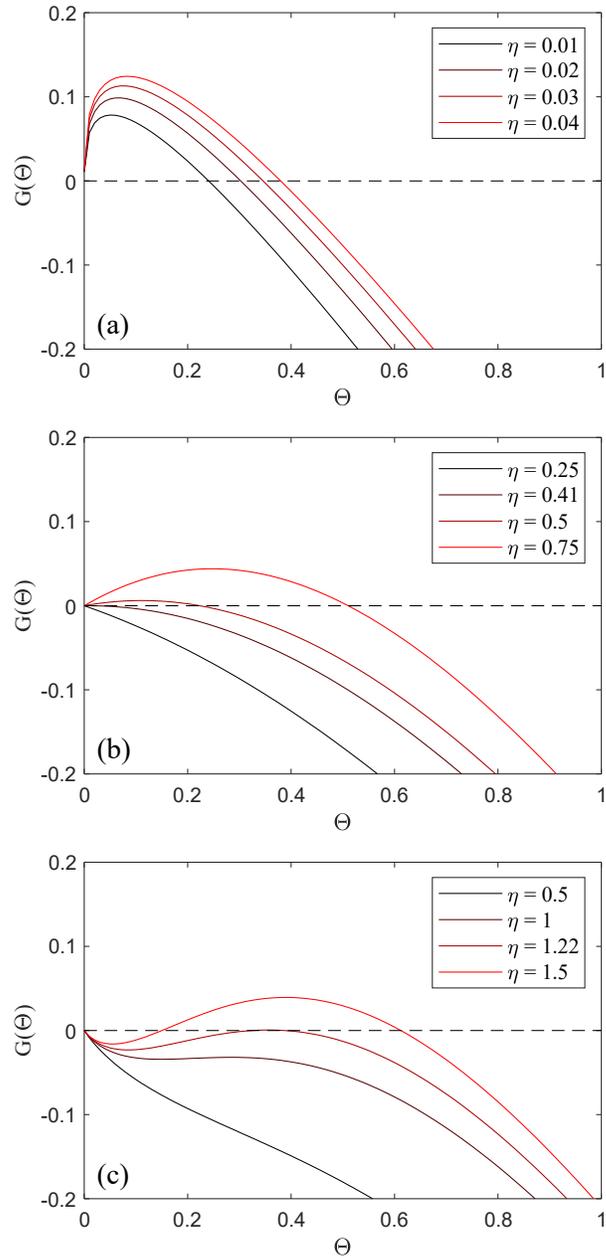}
\end{center}
\caption{Self-consistency function $G(\Theta)$ of SF 3-uniform hypergraphs with degree exponent (a) $\lambda=2.2$, (b) $2.5$, and (c) $2.8$, corresponding to cases i) $\lambda< \lambda_c$, ii) $\lambda=\lambda_c$, and iii) $\lambda> \lambda_c$ in the main text. The derivative of the function with respect to $\Theta$ at $\Theta=0$ (a) diverges, (b) is positive, and (c) is negative as $\Theta$ approaches zero.}
\label{self_consistency_equaiton}
\end{figure}

\begin{figure}
\includegraphics[width=1\textwidth]{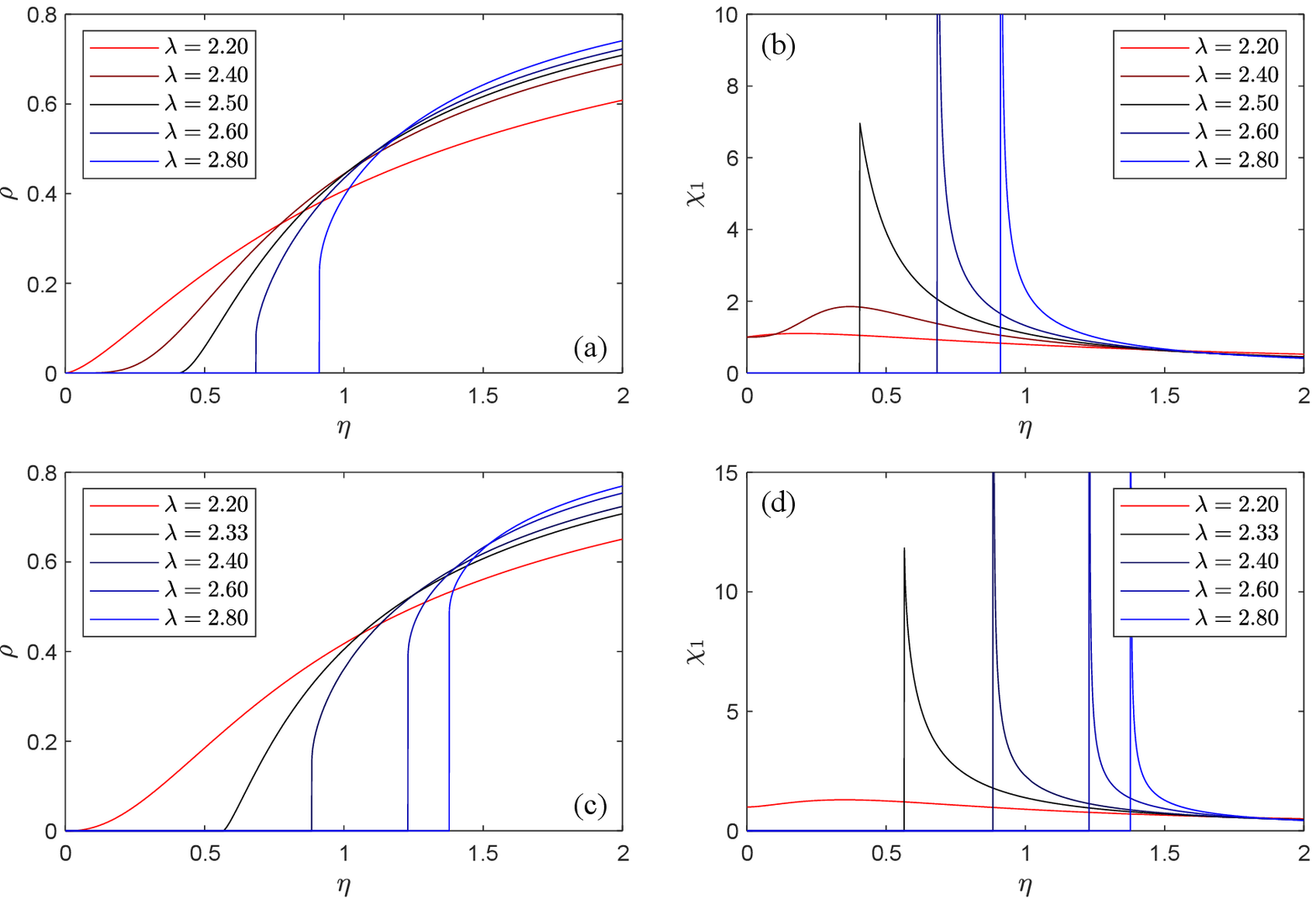}
\caption{Density of infected nodes versus control parameter $\lambda$ for various degree exponent values $\lambda$ for (a) $d=3$ and (c) $d=4$. Susceptibility versus control parameter $\lambda$ for various $\lambda$ values for (b) $d=3$ and (d) $d=4$. For $\lambda=2.2$ and $\lambda=2.4$, the transition point is $\lambda_c=0$, and for $\lambda=2.5, 2.6$, and $2.8$, $\lambda_c$ is finite. For $\lambda=2.2, 2.4$, and $2.8$, the transition is second-order, and for $\lambda=2.6$ and $2.8$, the transition is hybrid. For $\lambda\le \lambda_c$, the susceptibility converges to a finite value $1+d(d-2)$. For $\lambda>\lambda_{c}$, the susceptibility diverges as $\lambda \to \lambda_c^+$.}	
\label{mean_field_rho_chi}
\end{figure}

i) For $\lambda< \lambda_c$, $\lim_{\Theta\rightarrow0}G^{\prime}(\Theta)=\infty$. Because $G(0)=0$ and $G(1)<0$, there exists at least one solution $\Theta_{0} > 0$ for $\eta > 0$. Here, we find one such nontrivial stable solution $\Theta > 0$, leading to $\rho >0$. Therefore, a transition occurs at $\eta_c=0$. As $\eta$ is increased, both $\rho$ and $\Theta$ increase, and the transition is continuous. Analytically, we find that as $
\eta \to 0$, 
\begin{gather}
G(\Theta_0;k_m\eta)\simeq\frac{\left(\lambda-2\right)\pi}{\sin\left(\pi\lambda\right)}\left(k_m\eta\Theta_0^{d-1}\right)^{\lambda-2}-\Theta_0 = 0, \label{eq:gtheta_asymptotic_1}\\
\Theta_0\sim \eta^{\frac{\lambda-2}{1-(d-1)(\lambda-2)}}.
\end{gather}
The density of infection $\rho$ can also be calculated from Eq.~\eqref{eq:rho_hypergeom}:
\begin{equation}
\rho\sim \eta\Theta_0^{d-1}\sim \eta^{\frac{1}{1-(d-1)(\lambda-2)}}.
\label{eq:23}
\end{equation}
Thus, the exponent $\beta=1/[1-(d-1)(\lambda-2)]$. In particular, when $d=2$, $\rho\sim \eta^{1/(3-\lambda)}$~\cite{pastor-satorras2001a}.

ii) For $\lambda=\lambda_c$, the epidemic threshold is finite as $\eta_{c}=\frac{1}{k_m}\left[\frac{\sin\left(\pi/(d-1)\right)}{\pi/(d-1)}\right]^{d-1}$. Above $\eta_c$, $G^{\prime}(\Theta) > 0$, and thus there exists a finite $\Theta_0$ satisfying $G(\Theta_0)=0$. As $\eta \to \eta_{c}^+$, both $\rho$ and $\Theta_0$ decrease to zero. Thus, a second-order transition occurs at $\eta_c$. Specifically, the self-consistency function $G(\Theta)$ is written in Eq.~\eqref{eq:gtheta_asymptotic_1}. 
In this case, we need to consider higher-order terms of $G(\Theta)$ as 
\begin{align}
G(\Theta;k_m\eta)&\simeq\left[(\frac{\eta}{\eta_c})^{1/(d-1)}-1\right]\Theta-\frac{k_m\eta\Theta^{d-1}}{d-2}\label{eq:gtheta_asymptotic_2}\\
&\simeq \frac{1}{d-1}\left(\frac{\eta-\eta_c}{\eta_c}\right)\Theta-\frac{k_m\eta\Theta^{d-1}}{d-2}\,.\nonumber
\end{align} 
Therefore,
\begin{gather}
\Theta_0\sim\left(\eta-\eta_{c}\right)^{\frac{1}{d-2}}\,,\\
\rho\sim\left(\eta-\eta_{c}\right)^{\frac{d-1}{d-2}}\,.
\label{eq:26}
\end{gather}
Consequently, the critical exponent $\beta=(d-1)/(d-2)$ for $d > 2$. When $d=2$, $\rho\sim e^{-1/k_m\eta}$ was obtained~\cite{pastor-satorras2001a}.

iii) For $\lambda> \lambda_c$, $\lim_{\Theta\rightarrow0}G'(\Theta) <0$, and thus $\eta_{c}$ is finite. In this case, $\Theta_0$ and $\rho$ do not decrease to zero but are finite as $\eta \to \eta_c^+$. We calculate the asymptotic behaviors of $\Theta_0(\eta)-\Theta_0(\eta_{c})$ and $\rho(\eta)-\rho(\eta_{c})$. 
At the transition point, $G=0$ and $\partial_{\Theta} G=0$; thus, near this point,  
\begin{align}
G(\Theta;k_m\eta)&=\frac{1}{2}\frac{\partial^{2}G}{\partial\Theta^{2}}\left(\Delta\Theta\right)^{2}+\frac{\partial G}{\partial\eta}\Delta\eta+\cdots\,,\\ 
\Theta_0(\eta)-\Theta_0(\eta_{c})&\sim\left(\eta-\eta_{c}\right)^{1/2}\label{eq:theta}\,,\\
\rho(\eta)-\rho(\eta_{c})&\sim \left(\eta-\eta_{c}\right)^{1/2}\,,
\label{eq:29}
\end{align}
where $\Theta_0(\eta_c)$ and $\rho(\eta_c)$ are calculated using Eqs.~\eqref{eq:15} and~\eqref{eq:rho_hypergeom}, respectively.
Therefore, the transition is hybrid with the exponent $\beta=1/2$. 

\subsection{Susceptibility}
\label{sec:susceptibility}
The susceptibility is defined as the response of the order parameter, that is, the density of infection, to a conjugated field $h$:
\begin{equation}
\frac{d}{dt}\rho =-\rho+\eta\langle k \rangle \left(1-\rho\right)\Theta^{d-1}+(1-\rho)h\,.
\end{equation}
The conjugated field $h$ is implemented using the rate of spontaneous infection $S\rightarrow I$, i.e., the rate at which a susceptible node is changed to an infected state independently of the contagion process~\cite{lubeck}. The susceptibility is defined as the sensitivity of the density of infection to the conjugated field:
\begin{align}
\chi_1 & =\frac{\partial\rho}{\partial h}\,.
\end{align}
The differential equation for $\rho_k$ is written as 
\begin{equation}
\frac{d\rho_k}{dt}=-\rho_{k}+\eta k\left(1-\rho_{k}\right)\Theta^{d-1}+\left(1-\rho_{k}\right)h\,.
\end{equation}
The steady-state solution is obtained as 
\begin{equation}
\rho_{k}=\frac{h+\eta k\Theta^{d-1}}{1+h+\eta k\Theta^{d-1}}\,.
\end{equation}
The self-consistency equation is modified as follows: 
\begin{align}
G(\Theta,h)&={}_{2}F_{1}\left(\lambda-2,1;\lambda-1;-\frac{1+h}{k_m\eta\Theta^{d-1}}\right)\\ \nonumber
&+h\frac{\lambda-2}{\lambda-1}\frac{1}{k_m\eta\Theta^{d-1}}{}_{2}F_{1}\left(\lambda-1,1;\lambda;-\frac{1+h}{k_m\eta\Theta^{d-1}}\right)-\Theta\,.
\end{align}
The susceptibility is obtained using the following relation:  
\begin{equation}
\chi_1=\frac{\partial\rho}{\partial h}\Big|_{\eta,\Theta_0}-\frac{\partial\rho}{\partial\Theta}\Big|_{\eta,h}\frac{\partial G}{\partial h}\Big|_{\eta,\Theta_0}\left(\frac{\partial G}{\partial\Theta}\Big|_{\eta,h}\right)^{-1}\label{eq:chi_derivative}.
\end{equation}
Detailed calculations of the susceptibility are presented in~\ref{appendix:susceptibility}.

The results are as follows: 
i) For $\lambda\le \lambda_c$, the susceptibility converges to a finite value near the critical point, and therefore the critical exponent $\gamma_1 =0$.

ii) For $\lambda> \lambda_c$, the susceptibility diverges as $(\eta-\eta_c)^{-\gamma_1}$ with $\gamma_1=1/2$.

\subsection{Correlation size}

In the static model, the maximum degree diverges as $k_{\rm max}\sim N^{1/(\lambda-1)}$, which is called the natural cut-off~\cite{lee2006a}. We assign a weight $p_{i}$ to each node using Eq.~\eqref{eq:probability_node}. The exponent of the hyperedge degree distribution is $\lambda=1+{1}/{\mu}$.

\begin{figure}
\includegraphics[width=1\textwidth]{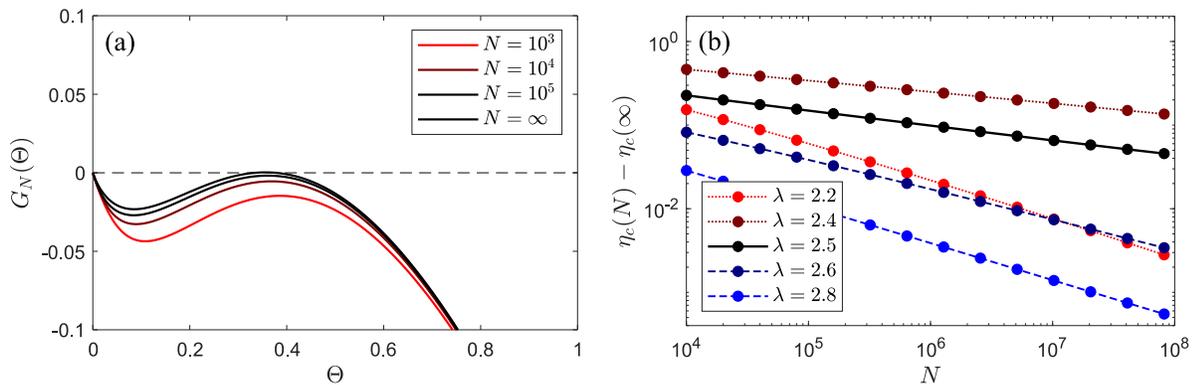}
\caption{
(a) Self-consistency function $G_N(\Theta)$ in finite systems versus $\Theta$ for 3-uniform hypergraphs with $\lambda=2.8$. (b) Deviation $\lambda_c(N)-\lambda_c(\infty)$ versus system size $N$ for various degree exponents $\lambda$. Red dotted lines denote $\lambda< \lambda_c= 2.5$; black solid lines denote $\lambda=\lambda_c$; and blue dashed lines do $\lambda > \lambda_c$.}
\label{gtheta_fsize}
\end{figure}

The self-consistency equation for finite systems reduces to 
\begin{align}
G_{N}(\Theta) 
&= \frac{1}{N\left<k\right>} \sum_{i=1}^N \frac{\eta \Theta^{d-1} k_i^2}{1 + \eta \Theta^{d-1} k_i} - \Theta,
\end{align}
where $k_i = \frac{N i^{-\mu}}{\sum_j j^{-\mu}}$. Further,
\begin{equation}
\frac{1}{N} \sum_{i=1}^N \frac{\eta \Theta^{d-1} k_i^2}{1 + \eta \Theta^{d-1} k_i} P(k_i) dk_i \simeq \int_{k_{\min}}^{k_{\max}} \frac{\eta \Theta^{d-1} k_i^2}{1 + \eta \Theta^{d-1} k_i} P(k_i) dk_i,
\end{equation}
where 
\begin{align}
k_{\min} = \left[\frac{1}{N} \sum_{j=1}^N \left(\frac{j}{N}\right)^{-\mu}\right]^{-1} 
&\simeq \int_0^1 x^{-\mu} dx - \int_0^{1/N} x^{-\mu} dx \\
&= \frac{\lambda-2}{\lambda-1} \left<k\right> \left(1-N^{-\frac{\lambda-2}{\lambda-1}}\right),
\end{align}
\begin{equation}
k_{\max} = k_{\min} N^{\frac{1}{\lambda-1}}.
\end{equation}
Therefore,
\begin{align}
G_{N}(\Theta) 
&\simeq \frac{\lambda-2}{k_m^{-\lambda+2}}\int_{k_{\min}}^{k_{\max}} dk\frac{\eta k^{-\lambda+2}\Theta^{d-1}}{1+\eta k\Theta^{d-1}} - \Theta \\
&\simeq G(\Theta) + k_m N^{-\frac{\lambda-2}{\lambda-1}} \frac{\eta k_m^{-\lambda+2}\Theta^{d-1}}{1+\eta k_m\Theta^{d-1}}  \nonumber \\
&\qquad - N^{-\frac{\lambda-2}{\lambda-1}} {}_{2}F_{1}\left(\lambda-2,1;\lambda-1;-\frac{1}{k_mN^{1/(\lambda-1)}\eta\Theta^{d-1}}\right),
\label{eq:gtheta_fsize}
\end{align}
where $G(\Theta)$ is the self-consistency function of the infinite system provided in Eq.~\eqref{eq:15}.
The solution of $G_N(\Theta)=0$ yields the density of infected nodes in finite systems. This function is illustrated in Fig. \ref{gtheta_fsize}(a) for a 3-uniform hypergraph with $\lambda= 2.8$. 

By expanding the finite-size self-consistency function in Eq.~\eqref{eq:gtheta_fsize} for large $N$, we can calculate the critical exponent of the correlation size, $\bar{\nu}$, which is defined by the relation $\eta_c(N)-\eta_c(\infty)\sim N^{-1/\bar{\nu}}$.

i) For $\lambda<\lambda_c$, $\lambda_c=0$, and thus $\lambda_c(N)$ is expected to be close to zero for large $N$. 
Therefore, for large $N$, 
\begin{equation}
N^{-(\lambda-2)/(\lambda-1)}{}_{2}F_{1}\left(\lambda-2,1;\lambda-1;-\frac{1}{k_mN^{1/(\lambda-1)}\lambda\Theta^{d-1}}\right)\simeq N^{-(\lambda-2)/(\lambda-1)}\label{eq:gtheta_correction_1},
\end{equation}
because the hypergeometric function converges rapidly to 1. 
The finite-size epidemic threshold is obtained when the maximum value of the function given by Eq.~\eqref{eq:gtheta_asymptotic_1} is equal to that given by Eq.~\eqref{eq:gtheta_correction_1}. 
Therefore, 
\begin{equation}
\eta_c(N)\sim N^{-\left[1-(d-1)(\lambda-2)\right]/(\lambda-1)}.
\label{eq:44}
\end{equation}
The inverse of the correlation size exponent is $1/\bar{\nu}=\left[1-(d-1)(\lambda-2)\right]/(\lambda-1)$, which approaches zero as $\lambda\rightarrow\lambda_{c}=2+{1}/(d-1)$. 

ii) For $\lambda=\lambda_c$, $(\lambda-\lambda_{c})\rightarrow0$, and $\Theta\rightarrow0$ with $\lambda_{c}>0$. 
The self-consistency function near the critical point is 
\begin{equation}
G_{N}(\Theta)=A\left(\lambda-\lambda_{c}\right)\Theta-B\Theta^{d-1}-N^{-(\lambda-2)/(\lambda-1)},
\end{equation}
where $A$ and $B$ are positive constants. 
Therefore,
\begin{equation}
\left( \eta-\eta_c \right)\sim N^{-\frac{d-2}{(d-1)^{2}(\lambda-1)}}.
\end{equation}
The inverse of the correlation size exponent becomes 
$1/\bar{\nu}=(d-2)/[(d-1)^{2}(\lambda-1)]$. 

iii) For $\lambda> \lambda_c$, the self-consistency function in finite systems becomes 
\begin{equation}
G_{N}(\Theta)=G(\Theta)+\frac{\partial G}{\partial\lambda}\left(\lambda-\lambda_{c}\right)-N^{-(\lambda-2)/(\lambda-1)}.
\end{equation}
Therefore,
\begin{equation}
\left(\eta-\eta_c\right)\sim N^{-\frac{\lambda-2}{(\lambda-1)}}.
\label{eq:48}
\end{equation}
The inverse of the correlation size exponent is $1/\bar{\nu}=(\lambda-2)/(\lambda-1)$. 

In this section, we obtained the critical exponents thorough the heterogeneous mean-field theory. The results are summarized in Tab.~\ref{tab:tab1}. Continuous (Discontinuous) transition occurs for $\lambda\le\lambda_c$ ($\lambda>\lambda_c$). At $\lambda=\lambda_c$, this is the boundary point where transition type and universality class are changed. Thus,  $\lambda=\lambda_c$ can be regarded as the tricritical point.

\begin{table*}[h!]
	\begin{center}
		\setlength{\tabcolsep}{8pt}
		{\renewcommand{\arraystretch}{1.3}
			\begin{tabular}{c c c c c c}
				\hline
				\hline
				~~$\lambda$~~ & $\eta_c$ & ~~$\rho_c$ & $\beta$ & $\gamma_1$ & $1/\bar{\nu}$  \\
				\hline
				$\lambda < \lambda_c$ & 0 & ~~0 & ~~$\frac{1}{1-(d-1)(\lambda-2)}$~~ &  0  & ~~$\frac{1-(d-1)(\lambda-2)}{\lambda-1}$~~   \\
				$\lambda = \lambda_c$ & finite & ~~0 & $\frac{d-1}{d-2}$ & 0 &  $\frac{d-2}{(d-1)^{2}(\lambda-1)}$ \\
				$\lambda > \lambda_c$ & finite & ~~~finite & $\frac{1}{2}$ & $\frac{1}{2}$ &  $\frac{\lambda-2}{\lambda-1}$ \\
				\hline
				\hline
		\end{tabular}}
	\end{center}
	\caption{Analytic solutions of the critical exponents for the $s$-SIS model.}
	\label{tab:tab1}
\end{table*}

\section{Numerical simulations}

\subsection{Numerical methods \label{subsec:numerical_methods}}

We perform numerical simulations using the sequential updating algorithm. The $s$-SIS model is simulated on an SF uniform hypergraph with $N$ nodes. Initially, all the nodes are assigned to fully infected states.
At each time step $t$, the following processes are applied:
\begin{itemize}
\item[i)] With probability $\kappa\equiv \eta/(1+\eta)$, we attempt the contagion process. We select a random hyperedge, and if the hyperedge satisfies the contagion condition, i.e., if all but one node of the hyperedge is in the infected state, the susceptible node in the hyperedge enters the infected state.
\item[ii)] With the remaining probability $1-\kappa=1/(1 + \eta)$, by contrast, we attempt the recovery process. A node is chosen at random, and if the chosen node is in the infected state, we change it to the susceptible state.
\item[iii)] If the number of active sites is zero, the simulation ends. Otherwise, the time $t$ is updated as $t \rightarrow t + 1/N$ in each step. Hereafter, we use the rescaled control parameter $\kappa$ instead of $\eta$.
\end{itemize}

A Markov process with an absorbing state in a finite-size system will ultimately reach the absorbing state. If the system has a nonzero probability of reaching the absorbing state after some time, the probability that the system remains active decreases exponentially and therefore converges to zero. To investigate the stationary state in a finite-size system in an absorbing state, samples surviving after a sufficiently long time are often taken as averages~\cite{marro2005}. This method is not computationally efficient, because the samples that have reached the absorbing state cannot be used to calculate the statistical properties of the stationary state. An alternative method is the quasistationary method~\cite{ferreira2011,ferreira2012}. In this method, if the system reaches an absorbing state, it reverts to an active configuration selected randomly from the history of the simulation. After a sufficiently long time, the system and the history simultaneously reach the stationary ensemble. In simulations, a list of $100$ previously visited configurations, is tracked and updated at each time step.

We performed the simulations in {\sl annealed hypergraphs}. An annealed hypergraph is a mean-field theoretical treatment of an ensemble of hypergraphs. We replaced the adjacency tensor with its ensemble average:
\begin{equation}
a_{\alpha}=\bar{a}_{\alpha}=f_{i_1 \cdots i_d}.
\end{equation}
The probability of a particular hyperedge $f_{i_1 \cdots i_d}$ in the static model of a uniform hypergraph was introduced in Sec.~\ref{sec:static_model}. For the probability of a hyperedge, we used $NK p_{i_1} \cdots p_{i_d}$, which is a valid approximation, even in the thermodynamic limit, as long as it is finite. This is a generalization of an annealed network. The annealed network, which was introduced as a randomly selected neighboring network~\cite{castellano2008}, has been widely used to study dynamical processes because heterogeneous mean-field theory and other mean-field theoretical approaches are exact in annealed networks~\cite{volz2008, pastor-satorras2015, lee2009, ferreira2011}.

\subsection{Numerical results}

\subsubsection{Static exponents\\}

\begin{figure}
\includegraphics[width=1.0\textwidth]{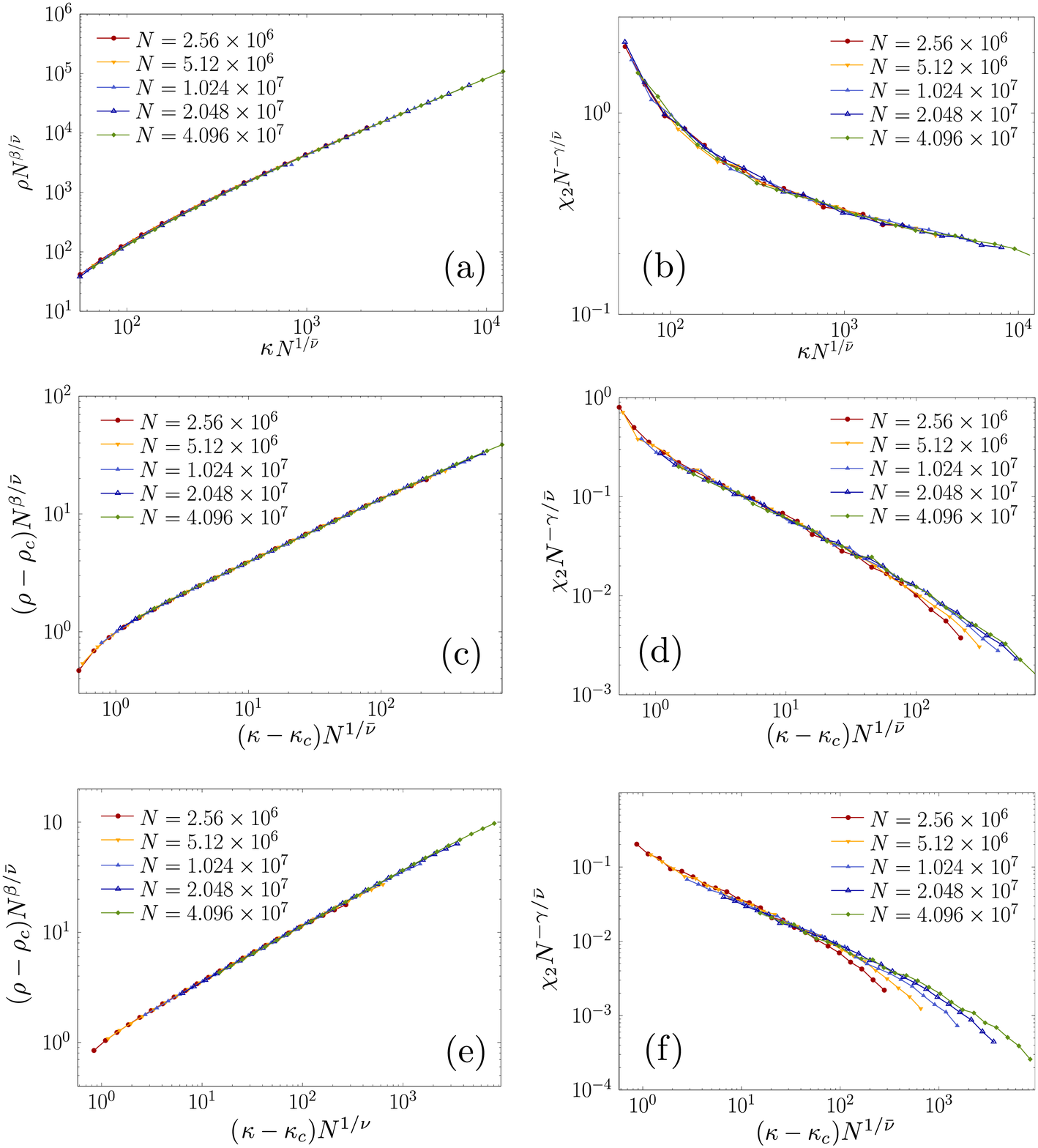}
\caption{Finite-size scaling analysis of the $s$-SIS model on SF $3$-uniform hypergraphs with three degree exponents: $\lambda=2.1<\lambda_c$ (a) and (b), $\lambda=2.9>\lambda_c$ (c) and (d), and $\lambda=3.5>\lambda_c$ (e) and (f). Scaling plots of $(\rho-\rho_c)N^{\beta/\bar{\nu}}$ versus $(\kappa-\kappa_c)N^{1/\bar{\nu}}$ are drawn, with (a) $\beta=1.25$ and $\bar{\nu}=1.59$, (c) $\beta=0.52$ and $\bar{\nu}=2.11$, and (e) $\beta=0.5$ and $\bar{\nu}=1.63$.
Scaling plots of $\chi_2N^{-\gamma_2/\bar{\nu}}$ versus $(\kappa-\kappa_c)N^{1/\bar{\nu}}$ are drawn, with (b) $\gamma_2=0.15$ and $\bar{\nu}=1.59$, (d) $\gamma_2=0.62$ and $\bar{\nu}=2.11$, and (f) $\gamma_2=0.62$ and $\bar{\nu}=1.63$.}
\label{fig:fig6}
\end{figure}

\begin{figure}
\includegraphics[width=1.0\textwidth]{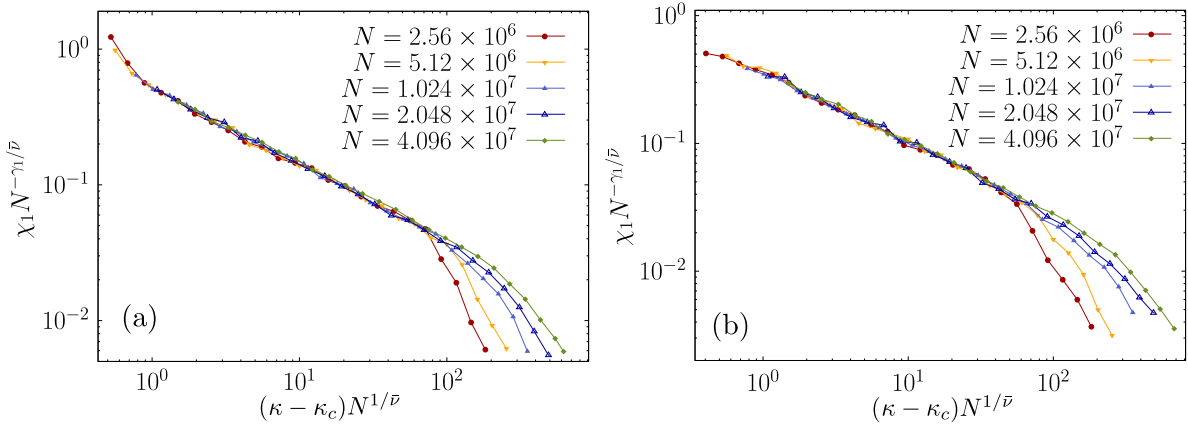}
\caption{Scaling plots of $\chi_1N^{-\gamma_1/\bar{\nu}}$ versus $(\kappa-\kappa_c)N^{1/\bar{\nu}}$ with degree exponents (a) $\lambda=2.9$ and (b) $\lambda=3.5$, with (a) $\gamma_1=0.48$ and $\bar{\nu}=2.11$, (b) $\gamma_1=0.50$ and $\bar{\nu}=1.63$.}
\label{fig:fig7}
\end{figure}

\begin{figure}
\includegraphics[width=1.0\textwidth]{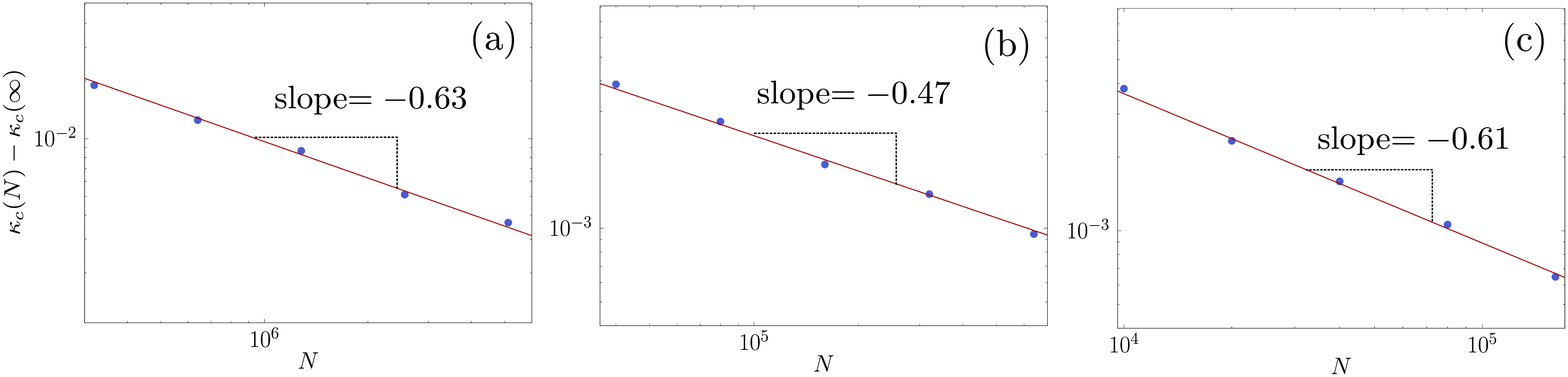}
\caption{Plots of $\kappa_c(N)-\kappa_c(\infty)$ versus $N$ on double-logarithmic scale for (a) $\lambda=2.1$, (b) $\lambda=2.9$, and (c) $\lambda=3.5$. Slope of each plot represents $-1/\bar{\nu}$.}
\label{fig:fig8}
\end{figure}

From Sec.~\ref{sec:analyticalResults}, the order parameter behaves as
\begin{equation}
\rho(\kappa)=\left\{
\begin{array}{lr}
0 & ~{\rm for}~~  \kappa < \kappa_c, \\
\rho_c+r(\kappa-\kappa_c)^{\beta} & ~{\rm for}~~ \kappa \ge \kappa_c, 
\end{array}
\right.
\label{eq:order}
\end{equation}
where $\rho_c$ is zero (finite) for $\lambda\le\lambda_c$ ($>\lambda_c$) and $\kappa_c$ is zero (finite) for $\lambda < \lambda_c$ ($\ge \lambda_c$) in the thermodynamic limit. Moreover, two types of susceptibilities are defined as follows: $\chi_1\equiv \partial \rho/\partial h\sim (\kappa-\kappa_c)^{-\gamma_1}$ and $\chi_2=N\,(\langle\rho^2\rangle - \langle\rho\rangle^2) /\langle\rho\rangle \sim (\kappa-\kappa_c)^{-\gamma_2}$. The correlation size exponent $\bar \nu$ is defined as $\kappa_c(N)-\kappa_c(\infty)\sim N^{-1/\bar{\nu}}$.

We performed simulations on a hypergraph with $d=3$ and the characteristic degree $\lambda_c=2.5$. Because the simulation results should be sensitive near $\lambda_c$, we chose $\lambda\in \{2.1, 2.9, 3.5\}$. We note that for the static model, a degree-degree correlation exists for $2 < \lambda <3$. Thus, the exponent $\bar{\nu}$ is expected to be different for $\lambda=2.9$ and $3.5$, whereas the other critical exponents, $\beta$ and $\gamma$, would be similar. Using finite-size scaling (FSS) analysis, we obtain the following: 

i) For $\lambda = 2.1 < \lambda_c$, we plot $\rho N^{\beta/\bar{\nu}}$ versus $\kappa N^{1/\bar{\nu}}$ for different system sizes but a fixed $d=3$ in Fig.~\ref{fig:fig6}(a). We find that the data points for different system sizes collapse onto a single curve for $\beta=1.25\pm 0.02$ and ${\bar{\nu}}=1.59\pm 0.01$. 
$\beta$ corresponds to the analytical result of Eq.~\eqref{eq:23}, but $\bar{\nu}$ is different with the analytical result of Eq.~\eqref{eq:44}. This discrepancy will be discussed in Sec.~\ref{sec:summary}.
For $\chi_2(\kappa)$, we plot $\chi_2 N^{-\gamma_2/\bar{\nu}}$ versus $(\kappa-\kappa_c)N^{1/\bar{\nu}}$ for $\gamma_2=0.15\pm 0.01$ and $\bar{\nu}=1.59$ in Fig.~\ref{fig:fig6}(b). Data points for systems of different sizes collapse well onto a single curve. 

ii) For $\lambda=2.9 > \lambda_c$, the transition point $\kappa_c$ and $\rho_c$ are numerically estimated to be $\approx 0.49462$ and $\approx 0.53877$, respectively, by solving the self-consistency equation [Eq.~\eqref{eq:15}] and using Eq.~\eqref{eq:rho_hypergeom}. On the basis of these values, we plot $(\rho-\rho_c) N^{\beta/\bar{\nu}}$ versus $(\kappa-\kappa_c) N^{1/\bar{\nu}}$ for $\beta=0.52\pm 0.02$ and $\bar{\nu}\approx 2.11\pm 0.01$ for different system sizes $N$ in Fig.~\ref{fig:fig6}(c). Thus, we confirm that the numerically estimated values are marginally consistent with the theoretical values from Eqs.~\eqref{eq:29} and \eqref{eq:48}. In Fig.~\ref{fig:fig6}(d), we plot the rescaled quantity $\chi_2 N^{-\gamma_2/\bar{\nu}}$ versus $(\kappa-\kappa_c)N^{1/\bar{\nu}}$ for different system sizes. We estimated $\gamma_2=0.62\pm 0.01$ and $\bar{\nu}=2.11$ using FSS analysis. Using the plot of $\chi_1 N^{-\gamma_1/\bar{\nu}}$ versus $(\kappa-\kappa_c)N^{1/\bar{\nu}}$ for different system sizes in Fig.~\ref{fig:fig7}, we estimated $\gamma_1=0.48\pm 0.02$. 

iii) For $\lambda=3.5$, we plot $(\rho-\rho_c) N^{\beta/\bar{\nu}}$ versus $(\kappa-\kappa_c) N^{1/\bar{\nu}}$ for different system sizes $N$ for $\beta=0.50\pm 0.01$ and ${\bar{\nu}}=1.63\pm 0.01$ in Fig.~\ref{fig:fig6}(e). 
For $\chi_2(\kappa)$, we plot $\chi_2 N^{-\gamma_2/\bar{\nu}}$ versus $(\kappa-\kappa_c)N^{1/\bar{\nu}}$ for $\gamma_2=0.62\pm 0.01$ and ${\bar{\nu}}=1.63$. The data collapse well onto a single curve, as shown in Fig.~\ref{fig:fig6}(f). We plot $\chi_1 N^{-\gamma_1/\bar{\nu}}$ versus $(\kappa-\kappa_c)N^{1/\bar{\nu}}$ in Fig.~\ref{fig:fig7}. We estimated $\gamma_1=0.50\pm 0.02$. The obtained values, $\beta=0.5\pm 0.01$, $\gamma_2=0.62\pm 0.01$, and $\bar{\nu}=1.63\pm 0.01$, marginally satisfy the hyperscaling relation $\bar{\nu}=2\beta+\gamma_2$. 

The correlation size exponent is measured directly as $\kappa_c(N)-\kappa_c(\infty)\sim N^{-1/\bar{\nu}}$ with $1/\bar{\nu}=0.63$, $0.47$, and $0.61$ in Fig.~\ref{fig:fig8}, which correspond to $\bar{\nu}\simeq 1.59$, $2.13$, and $1.64$ for $\lambda=2.1$, $2.9$, and $3.5$, respectively. These values are in reasonably good agreement with the values $\bar{\nu}=1,59 \pm 0.01, 2.11 \pm 0.01$, and $1.63\pm 0.01$ obtained by FSS analysis in Fig.~\ref{fig:fig6}. We summarize the numerical values in Table~\ref{tab:tab2}.

\begin{table*}[h!]
	\begin{center}
		\setlength{\tabcolsep}{8pt}
		{\renewcommand{\arraystretch}{1.3}
			\begin{tabular}{ccc cc cc}
				\hline
				\hline
				$\lambda$ & $\kappa_c$ & $\rho_c$& $\beta$ & $\gamma_1$ & $\gamma_2$ & $\bar{\nu}$  \\
				\hline
				\multirow{2}{*}{$2.1$} & $0$ & $0$ & $1.25\pm 0.02$& $0$ & $0.15\pm 0.01$& $1.59\pm 0.01$   \\
				&  &  & $(1.25)$& ($0$) & & $(1.35)$   \\\hline
				\multirow{2}{*}{$2.9$} & $0.49462$ & $0.268306$ & $0.52\pm 0.02$& $0.48\pm 0.02$& $0.62\pm 0.01$ & $2.11\pm 0.01$ \\
				&  &  & $(0.50)$& ($0.50$) & & $(2.11)$   \\\hline
				\multirow{2}{*}{$3.5$} & $0.53877$ & $0.395602$& $0.50\pm 0.01$& $0.50\pm 0.02$&   $0.62\pm 0.01$ & $1.63\pm 0.01$ \\
				&  &  & $(0.50)$& ($0.50$) & & $(1.67)$   \\
				\hline
				\hline
		\end{tabular}}
		\caption{Numerical list of critical exponents of the $s$-SIS model obtained by the FSS method. Theoretical values calculated in Sec.~\ref{sec:analyticalResults} are presented in parentheses.}
		\label{tab:tab2}
	\end{center}
\end{table*}

\begin{figure}
\includegraphics[width=1\textwidth]{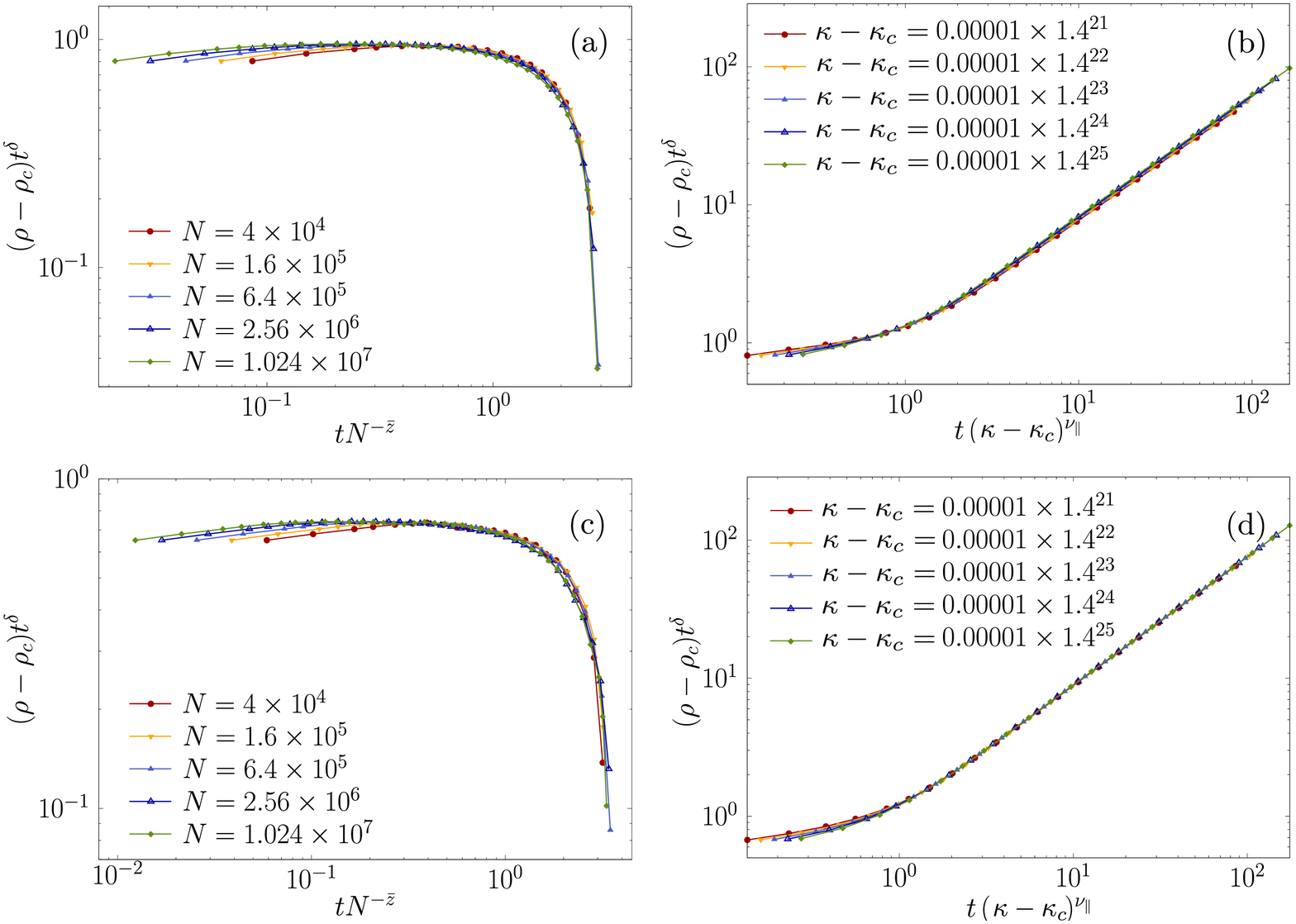}
\caption{Scaling plots of the density of infection $\rho(t)$ starting from the fully infected state versus $tN^{-\bar z}$ (a) and (c) and $t(\kappa-\kappa_c)^{\nu_{\|}}$ (b) and (d) for $\lambda=2.9$ (a) and (b) and $\lambda=3.5$ (c) and (d). The dynamical critical exponents $\delta=0.89$, $\bar{z}=0.26$, and $\nu_{\|}=0.56$ are obtained from (a) and (b), and $\delta=0.86$, $\bar{z}=0.32$, and $\nu_{\|}=0.53$ are obtained from (c) and (d).}
\label{fig:fig9}
\end{figure}

\subsubsection{Dynamic exponents\\}

Next, we also performed dynamical FSS analysis to obtain the dynamic exponents. 
We consider the temporal dynamics of the density of infection starting from a fully infected state. The average density of infection at time $t$ over many realizations, $\rho(t)$, shows critical behavior when the contagion rate is equal to the critical value $\kappa_c$. We choose $\lambda\in \{2.9, 3.5\}$ because for $\lambda<\lambda_c$, the critical point $\kappa_c$ becomes zero, and only a decay process remains.
In this section, we change the notation of $\bar{\nu}$ to $\bar{\nu}_{\bot}$ as a counterpart of the mean survival time exponent $\nu_{\|}$.

i) For $\lambda=2.9$, we plot $(\rho-\rho_c) t^{\delta}$ versus $t N^{-\bar{z}}$ for different system sizes $N$ in Fig.~\ref{fig:fig9}(a). 
Here, the dynamical critical exponents are defined conventionally as $\delta=\beta/\nu_{\|}$ and $\bar{z}\equiv \nu_{\|}/\bar{\nu}=\nu_{\|}/d\nu_{\bot}$.
In Fig.~\ref{fig:fig9}(b), we plot the rescaled quantity $(\rho-\rho_c) t^{\delta}$ versus $t(\kappa-\kappa_c)N^{\nu_{\|}}$.
$\nu_{\|}$ is the mean survival time exponent associated with the relaxation time.
We estimated the dynamical critical exponents as $\delta=0.89\pm 0.02$, $\bar{z}=0.26 \pm 0.01$, and $\nu_{\|}=0.56\pm 0.01$. 

ii) For $\lambda=3.5$, we used a method similar to that used in i).
We estimated the dynamical critical exponents as $\delta=0.93\pm 0.02$, $\bar{z}=0.32 \pm 0.01$, and $\nu_{\|}=0.53\pm 0.01$. 

The critical exponents $\{\delta,\, \bar{z},\, \nu_{\|}\}$ obtained using dynamical FSS and the $\{\beta,\, \bar{\nu},\, \gamma_2\}$ values obtained using steady-state FSS are comparable.

\begin{table*}[h!]
\begin{center}
\setlength{\tabcolsep}{8pt}
{\renewcommand{\arraystretch}{1.3}
\begin{tabular}{cccc}
    \hline
    \hline
    $\lambda$ & $\delta$ & $\bar{z}$ & $\nu_{\|}$ \\
   \hline
  $2.9$ & $0.89\pm 0.02$ & $0.26\pm 0.01$ & $0.56\pm 0.01$\\
  $3.5$ &$0.93\pm 0.02$ & $0.32\pm 0.01$ & $0.53\pm 0.01$ \\
  \hline
    \hline
\end{tabular}}
\caption{Dynamic critical exponents of $s$-SIS model obtained using the dynamical FSS method.}
\label{tab:tab3}
\end{center}
\end{table*}
\section{Summary}
\label{sec:summary}

In summary, we investigated the phase transitions and critical phenomena of the $s$-SIS model in SF uniform hypergraphs. We proposed a static model of the uniform hypergraph, which is a generalization of the static model of a complex network. We showed that the model indeed exhibits a degree distribution with a power-law tail.

Using the heterogeneous mean-field theory, we analytically studied the $s$-SIS model. We showed that the system exhibits rich phase transition and critical phenomena when the exponent of the degree distribution $\lambda$ is larger than two. There exists a characteristic degree $\lambda_c=2+1/(d-2)$. For $\lambda<\lambda_c$, the epidemic threshold vanishes. Thus, there exists a stationary state for an arbitrarily small contagion rate in the thermodynamic limit. The susceptibility $\chi_2$, the fluctuations of the order parameter, diverges as $\kappa \to 0$. Thus, a second-order contagion transition occurs at $\kappa_c=0$. For $\lambda=\lambda_c$, the epidemic threshold
becomes finite and the susceptibility $\chi_2$ diverges as $\kappa\to\kappa_c$. Thus, a second-order contagion transition occurs. For $\lambda>\lambda_c$, the system undergoes a hybrid phase transition at a finite transition point $\kappa_c$. The susceptibility diverges at the transition point. We note that in a previous study~\cite{iacopini2019}, a discontinuous contagion transition was observed owing to higher-order interactions in a different model; however, we observed a hybrid phase transition, which exhibits a discontinuous transition with criticality at the same transition point. We also notice that for the static model, when the degree exponent is $2 < \lambda \le 3$, a degree-degree correlation exists. Consequently, the correlation size exponent $\bar{\nu}_\bot$ differs from that for $\lambda > 3$. Accordingly, 
whereas the measured critical exponents $\beta$ and $\gamma$ are close to each other for $\lambda_c < \lambda < 3$ and $\lambda > 3$, the dynamic exponents $\delta$ and $\bar z$ associated with $\bar\nu_\bot$ and $\nu_{\|}$ are different.   

We performed numerical simulations of annealed SF $3$-uniform hypergraphs with $\lambda_c=2.5$ and the degree exponents $\lambda=2.1$, $2.9$, and $3.5$. Using dynamical FSS and steady-state FSS, the critical exponents $\{\delta,\, \bar{z},\, \nu_{\|}\}$ and $\{\beta,\, \bar{\nu}_\bot,\, \gamma_1,\, \gamma_2\}$ are listed in Tables~\ref{tab:tab2} and \ref{tab:tab3}, respectively. The two methods are consistent within the error bars. Finally, the numerical values of the critical exponents $\{\beta,\, \bar{\nu}_\bot,\, \gamma_2\}$ are consistent with the theoretical values based on the heterogeneous mean-field theory in Sec.~\ref{sec:analyticalResults}. They are listed in Table~\ref{tab:tab1}.

\ack
This research was supported by the NRF, Grant No.~NRF-2014R1A3A2069005 (BK).
\appendix 

\section{Degree distribution of static model}\label{appendix:degree_dist} 

Throughout this construction algorithm, a node is selected with probability $1 - \left( 1-p_i \right)^d \simeq d p_i$. Therefore, the probability that a node $i$ has degree $k$ follows the Poisson distribution:
$P_i^{(R)} (k) = {\left\langle k_i \right\rangle}^k \exp \left( -\left\langle k_i \right\rangle \right) /k!$. 
The degree distribution is then
\begin{eqnarray}
P^{(R)}(k) 
&= \frac{1}{N}\sum P_{i}(k)
\simeq \int_{{k}_{\rm{min}}}^{{k}_{\rm max}}d\langle k_{i}\rangle P\left(\langle k_{i}\rangle \right)\frac{\left\langle k_{i}\right\rangle ^{k}\exp\left(-\left\langle k_{i}\right\rangle \right)}{k!}\\
&= \frac{\left(\lambda-1\right)}{\left\langle k_{i}\right\rangle _{\mathrm{min}}^{-\lambda+1}-\left\langle k_{i}\right\rangle _{\max}^{-\lambda+1} } \frac{1}{k!} \int_{\left\langle k_{i}\right\rangle _{\mathrm{min}}}^{\left\langle k_{i}\right\rangle _{\max}}d\left\langle k_{i}\right\rangle \left\langle k_{i}\right\rangle ^{-\lambda+k}\exp\left(-\left\langle k_{i}\right\rangle \right).
\end{eqnarray}
In the thermodynamic limit, 
$\left\langle k_{i}\right\rangle _{\max}\rightarrow\infty$ and $\langle k_{i}\rangle _{\min}\rightarrow\frac{\lambda-2}{\lambda-1}\left\langle k\right\rangle$. Further,
\begin{equation}
\lim_{N\rightarrow\infty}P^{(R)}(k) = \left(\lambda-1\right) k_m^{\lambda-1} \frac{\Gamma\left(-\lambda+k+1,k_m\right)}{\Gamma(k+1)}\sim k^{-\lambda}
\end{equation}
for sufficiently large $k$. 
Therefore, the tail of the degree distribution of a static model of a uniform hypergraph follows a power law.

\section{Asymptotic behavior of $G^\prime(\Theta)$} 
\label{appendix:gtheta_derivative}

Using the identity
\begin{eqnarray}
_{2}F_{1}\left(a,b;c;-z\right)&=&\frac{z^{-a}\Gamma(c)\Gamma(b-a){}_{2}F_{1}\left(a,a-c+1;a-b+1;-\frac{1}{z}\right)}{\Gamma(b)\Gamma(c-a)}\\ \nonumber
&+&\frac{z^{-b}\Gamma(c)\Gamma(a-b){}_{2}F_{1}\left(b,b-c+1;-a+b+1;-\frac{1}{z}\right)}{\Gamma(a)\Gamma(c-b)}\label{eq:hypergeom_identity},
\end{eqnarray}
we can obtain the asymptotic behavior of the hypergeometric function $_{2}F_{1}\left(a,b;c;-z\right)$ as $z\rightarrow\infty$:
\begin{equation}
_{2}F_{1}\left(a,b;c;-z\right)\sim\begin{cases}
\frac{\Gamma(c)\Gamma(b-a)}{\Gamma(b)\Gamma(c-a)}z^{-a} & a<b\\
\frac{\Gamma(c)\Gamma(a-b)}{\Gamma(a)\Gamma(c-b)}z^{-b} & a>b
\end{cases}.
\end{equation}
The formula also allows us to calculate the next dominant terms proportional to $z^{-a-1}$, $z^{-a-2}$, $\cdots$ and $z^{-b-1}$, $z^{-b-2}$, $\cdots$. 
As $\Theta\rightarrow0$,
\begin{equation}
G^{\prime}(\Theta)\sim\begin{cases}
\frac{\pi(d-1)(\lambda-2)^{2}}{\sin\left(\pi\lambda\right)}\left(k_m\lambda\right)^{\lambda-2}\Theta^{(d-1)\lambda-(d-1)-d}-1 & \lambda<3\\
\frac{(d-1)(\lambda-2)}{(\lambda-3)}k_m\lambda\Theta^{d-2}-1 & \lambda>3
\end{cases}.
\end{equation}
Then we obtain Eq.~\eqref{eq:gtheta_derivative_asymptotic}.

\section{Susceptibility}
\label{appendix:susceptibility}
To calculate Eq.~\eqref{eq:chi_derivative}, we first take the derivatives and then set $h=0$ and $\Theta=\Theta_0$: 
\begin{align}
\frac{\partial\rho}{\partial h}\Big|_{\eta,\Theta}&=1-{}_{2}F_{1}\left(\lambda-1,1;\lambda;-\frac{1}{k_{m}\eta\Theta_0^{d-1}}\right) \label{eq:c1}\\
&-\frac{\lambda-1}{\lambda}\frac{1}{k_{m}\eta\Theta_0^{d-1}}{}_{2}F_{1}\left(\lambda,2;\lambda+1;-\frac{1}{k_{m}\eta\Theta_0^{d-1}}\right) \,,\nonumber\\ 
\frac{\partial\rho}{\partial\Theta}\Big|_{\eta,h}&=\frac{(d-1)(\lambda-1)}{\lambda}\frac{1}{k_{m}\eta\Theta_0^{d}}{}_{2}F_{1}\left(\lambda,2;\lambda+1;-\frac{1}{k_{m}\eta\Theta_0^{d-1}}\right)\label{eq:c2}\,,\\
\frac{\partial G}{\partial h}\Big|_{\eta,\Theta}&=\frac{\lambda-2}{\lambda-1}\frac{1}{k_{m}\eta\Theta_0^{d-1}}\Bigg[{}_{2}F_{1}\left(\lambda-1,1;\lambda;-\frac{1}{k_{m}\eta\Theta_0^{d-1}}\right)\label{eq:c3}\\
&-{}_{2}F_{1}\left(\lambda-1,2;\lambda;-\frac{1}{k_{m}\eta\Theta_0^{d-1}}\right)\Bigg]\,,\nonumber\\
\frac{\partial G}{\partial\Theta}\Big|_{\eta,h}&=\frac{(d-1)(\lambda-2)}{\lambda-1}\frac{1}{k_{m}\eta\Theta_0^{d}}{}_{2}F_{1}\left(\lambda-1,2;\lambda;-\frac{1}{k_{m}\eta\Theta_0^{d-1}}\right)-1\label{eq:c4} \,.
\end{align}
Using Eq.~\eqref{eq:hypergeometric_taylor}, we obtain the following:

i) For $\lambda<\lambda_c$, Eq.~\eqref{eq:c1} becomes $1$, and all other terms vanish in the limit $\Theta_0\rightarrow0$ and $\eta\rightarrow0$.
Therefore, $\chi_1=1$ near the critical point, and the critical exponent of the susceptibility, $\gamma_1$, is zero.

ii) For $\lambda=\lambda_c$, Eqs.~\eqref{eq:c1}--\eqref{eq:c4} in the limit $\Theta_0\rightarrow0$ and $\eta\rightarrow \eta_c$ are given as
\begin{align}
\frac{\partial\rho}{\partial h}\Big|_{\eta,\Theta_0}&=1 \,,\qquad
\frac{\partial\rho}{\partial\Theta}\Big|_{\eta,h}\sim d(d-2)\frac{\eta-\eta_c}{\eta_c}\,,\\
\frac{\partial G}{\partial h}\Big|_{\eta,\Theta_0}&\sim 1\,,\qquad
\frac{\partial G}{\partial\Theta}\Big|_{\eta,h}\sim -\frac{d-2}{d-1}\frac{\eta-\eta_c}{\eta_c} \,.
\end{align}
The susceptibility is given by $\chi_1\sim 1+d(d-1)$.

iii) For $\lambda > \lambda_c$, Eq.~\eqref{eq:c4} exhibits singular behavior, and Eqs.~\eqref{eq:c1}--\eqref{eq:c3} are finite. Hence, the susceptibility diverges near the critical point. Eq.~\eqref{eq:c4} is calculated as
\begin{align}
\frac{\partial G}{\partial\Theta}\sim\frac{\partial^{2}G}{\partial\Theta^{2}}\left(\Delta\Theta_0\right)\label{eq:c7}\,.
\end{align}
Inserting Eq.~\eqref{eq:theta} into Eq.~\eqref{eq:c7} yields $\chi_1\sim(\eta-\eta_{c})^{-1/2}$, and therefore $\gamma_1=1/2$.

\end{document}